\documentclass[aps,prd,10pt,twocolumn,nofootinbib]{revtex4}

\usepackage{epsfig}
\usepackage{bm}
\usepackage{latexsym}
\usepackage{natbib}
\usepackage{url}
\usepackage{dcolumn}
\usepackage{color}
\usepackage{amsfonts,amssymb,amsmath}
\usepackage{graphicx,epsfig}
\usepackage{psfrag}
\usepackage{subfigure}
\usepackage{hyperref}
\hypersetup{colorlinks=true}
\usepackage{mathtools}
\usepackage{enumitem}
\usepackage{float}
\usepackage{mathrsfs}
\usepackage[dvipsnames]{xcolor}
\usepackage[compat=1.1.0]{tikz-feynman}
\hypersetup{
  colorlinks,
  citecolor=red,
  linkcolor=red,
  urlcolor=blue}

\usepackage{comment}

\begin{document}

\title{Synergy Between Hubble Tension Motivated Self-Interacting Neutrino and KeV-Sterile Neutrino Dark Matter}

\author{
Mansi Dhuria$^{a,}$\footnote{Mansi.dhuria@sot.pdpu.ac.in}, Abinas Pradhan$^{b,}$\footnote{abinaspradhan93@gmail.com}
}

\affiliation{
$^a$ Department of Physics, School of Energy Technology, Pandit Deendayal Energy University (PDEU), Gandhinagar-382426, Gujarat, India \\
$^b$ Basic Sciences Department, Institute of Infrastructure, Technology, Research and Management (IITRAM), Ahmedabad-380026, Gujarat, India
}

\begin{abstract}
\noindent The discrepancy between the value of Hubble constant measured by CMB observations and local low-redshift based observations has proposed many solutions which require the existence of Physics beyond Standard Model (SM). One of the interesting solutions is based on considering the strong self-interaction between Standard Model (SM) neutrinos through an additional scalar/vector mediator. Interestingly, the strong self-interaction between SM neutrinos also play an important role in obtaining KeV-sterile neutrino as a viable Dark Matter (DM) candidate through the famous Dodelson-Widrow mechanism. In this work, we have tried to find the synergy between the parameter space of active-sterile neutrino mixing vs mass of sterile neutrino allowed by Hubble tension solution and the requirement of getting KeV-sterile neutrino as DM candidate. Interestingly, we get a large amount of parameter space that is consistent with both the requirements and also free from X-Ray constraints. Finally, we have embedded this scenario in a consistent supersymmetric model of particle physics. In this framework, we have shown that the value of sterile neutrino mass, SM neutrino mass and the required mixing angle can be naturally obtained by considering the supersymmetry breaking scale to be around ${\cal O}(10)$ TeV. Thus, it would  give an interesting testing ground for supersymmetry as well as signatures of Warm Dark Matter (WDM).
\end{abstract}

\maketitle

\section{Introduction}
The accumulation of tension between the value of Hubble constant obtained from recent Cosmic Microwave Background (CMB) measurements and the local distance ladder measurements has indicated the need to look for new physics beyond Standard $\Lambda$CDM model~\cite{DiValentino:2021izs,Abdalla:2022yfr,Vagnozzi:2019ezj}. In particular, the recent CMB measurements gives the value of present day Hubble constant  $H_0 = 67.27\pm 0.60$ km/s/Mpc$^{-1}$~\cite{Planck:2018vyg} while the analysis based on Cepheid-calibrated Type Ia supernovae by the SH0ES collaboration gives $H_0 =74.03 \pm 1.42$ km/s/Mpc$^{-1}$~\cite{Riess:2019cxk,riess2018new}.
While the tension might exist due to issues in systematic, several solutions have been proposed to address this by proposing modification either in the early universe or the late-universe physics~\cite{DiValentino:2021izs,Abdalla:2022yfr,Vagnozzi:2019ezj,Niedermann:2021ijp,Niedermann:2021vgd,Rezazadeh:2022lsf,Dainotti:2022bzg,Dainotti:2021pqg}.
All these solutions are based on obtaining the high value of $H_0$ from CMB based measurements.
One interesting solution in the context of non-$\Lambda$CDM model was proposed in \cite{Cyr-Racine:2013jua} by invoking  new non-standard Fermi-like four-fermion interaction of massless neutrinos with each other, parameterized by an effective coupling constant $G_{\rm eff}$. It has been argued in the literature that the strong self-interactions between neutrinos might delay the onset of neutrino free-streaming until close to the onset of matter-radiation equality, thus leading to a higher value of $H_0$. The constraints on the strength of $G_{\rm eff}$ have been studied in \cite{Lancaster:2017ksf,Oldengott:2017fhy,Huang:2017egl,Forastieri:2019cuf,RoyChoudhury:2020dmd} by assuming the non-renormalizable interaction mediated through heavy mediators (with mass greater than neutrino decoupling temperature) as well as light mediators. The subsequent studies in this direction have also taken into account the effect on these parameters by considering self-interactions between specific flavours of neutrino~\cite{Das:2021guu}. Overall, it has been found  that the fit to CMB data prefers a specific range of the self-interaction strength between neutrinos for all three flavors of  neutrino. However, a large amount of the range of required $G_{\rm eff}$ has been ruled out by various laboratory and cosmological bounds~\cite{Blinov:2019gcj,Lyu:2020lps,RoyChoudhury:2022rva}.

As the self-interactions between neutrinos are theoretically possible, they have also been explored to analyse the possibility of KeV-sterile neutrino as a viable WDM candidate by proposing modified Dodelson-Widrow (DW) mechanism induced through non-standard neutrino self interactions~\cite{DeGouvea:2019wpf,Kelly:2020pcy,Benso_2022}. The standard DW mechanism allows the production of sterile neutrino DM for an appropriate values of active-sterile neutrino mixing angle~\cite{Dodelson:1993je} and around KeV-scale mass. However,  the non-zero mixing also allows the sterile neutrino to decay slowly into photon and active neutrino at one-loop level with decay width~\cite{Abazajian:2021zui,Dekker:2021bos}
\begin{equation}
\Gamma_{\nu_s}(m_{\nu_s},\theta) = 1.38 \times 10^{-29} {\rm s^{-1}} \Biggl[\frac{\sin^2{2\theta}}{10^{-7}}\Biggr]\Biggl[\frac{m_{\nu_s}}{1~{\rm KeV}}\Biggr],\nonumber
\end{equation}
where $\theta$ is the mixing angle and $m_{\nu_s}$ is sterile neutrino mass. This will produce monochromatic X-Ray photons with energy $E_{\gamma}=m_{\nu_s}/2$, which has been observed by various X-Ray telescopes. As relic abundance produced through standard DW mechanism can be recasted into the parameter space of $m_{\nu_s}-\sin^2{2\theta}$, it has been shown that the entire parameter space of $m_{\nu_s}-\sin^2{2\theta}$ consistent with sterile neutrino DM has been ruled out by X-Ray observations and the observation of 3.5 KeV X-Ray line~\cite{Tremaine:1979we, Boyarsky:2008ju, Merle:2015vzu, Abazajian:2017tcc, Watson:2011dw, Horiuchi:2013noa, Perez:2016tcq, Dessert:2018qih, Ng:2019gch, calore2022constraints}. However, it has been discussed in \cite{DeGouvea:2019wpf,Kelly:2020pcy,Benso_2022} that the DW mechanism can be modified by taking into account the production of active neutrinos through self-interaction. Interestingly, the modified version of the same would allow the right value of relic abundance of KeV-sterile neutrino for the range of active-sterile neutrino mixing angle yet unconstrained by X-Ray observations~\cite{DeGouvea:2019wpf,Kelly:2020pcy,Benso_2022}.  Another interesting approach to avoid X-Ray bounds without considering neutrino non-standard self-interactions is studied in~\cite{Benso_2019}.

The motivation to consider Cold Dark Matter (CDM) as a viable DM candidate is being faded due to non-observation of signatures of well-motivated CDM candidate, neutralino,  at the Large Hardon Collider (LHC). Additionally, though the existence of CDM is  totally compatible with the observations of large scale structure of the universe, it is inconsistent with the  observed structure at small scale structure of the universe~\cite{Drewes:2016upu}. On the other hand, WDM candidates have been able to successfully address almost all the effects found at small scale structure of the universe.  In view of this, there is a growing interest in exploring the viability of popular WDM candidates such as sterile neutrino, axions etc.~\cite{Drewes:2016upu,Viel:2013fqw}. As the self-interacting neutrino solution of Hubble tension indicates  new physics beyond $\Lambda$CDM model, it would be worthwhile to explore if there is a synergy between the self-interacting neutrino model required by Hubble tension and  modified DW-based production mechanism of KeV-neutrino DM. As the cosmological data is becoming more and more precise, the intimate connection between the two issues shall open up a new window to the observational signatures of DM from recent cosmological data. Thus, in this work,  we have calculated the relic abundance of KeV-sterile neutrino by taking a specific range of the strength of self-interactions required to solve Hubble tension. We have also obtained the $m_{\nu_s}-\sin^2{2\theta}$ parameter space for the preferred range of $G_{\rm eff}$ and analysed if it is still consistent with constraints from X-Ray observations. 

The plan of the rest of the paper is as follows: In \$II, we briefly review the physics of self-interacting neutrino and its role in alleviating the Hubble tension. In \$III.A and \$III.B, we review the calculation of relic abundance of sterile neutrino produced through standard DW mechanism and the modified DW mechanism respectively. In \$IV, we scanned over the parameter space of neutrino mass and mixing between sterile and SM neutrino suited to obtain the right value of relic abundance of KeV-sterile neutrino for the range of $G_{\rm eff}$ obtained from the consideration of Hubble tension.  In \$V, we present a toy version of the phenomenological model which would naturally explain the required mass of SM and sterile neutrino DM while keeping suppressed value of active-sterile neutrino mixing angle as required from the results obtained in \$IV. In subsection \$V.a, we verify that the decay of massive scalar mediator $\phi$ into sterile neutrinos would not thermalize the right-handed neutrino before the epoch of Big-Bang Nucleosynthesis (BBN). Hence, there  would not be any effect  in the observed relativistic degrees of freedom $\Delta N_{\rm eff}$. Finally, in \$VI, we discuss our results with interesting conclusions and future directions.  
\section{Self-interacting neutrino and the Hubble Tension}
In this section, we briefly review the effect of self-interacting nature of neutrinos on the CMB power spectrum which leads to the change in the present day Hubble constant. 
In the standard $\Lambda$CDM model, the perturbations of active neutrinos free-streaming through the photon-baryon plasma generates the anisotropic stress which further modifies the gravitational potential and photon perturbations~\cite{Das:2020xke, Ghosh:2019tab}.
Given that the neutrino travels  nearly at the speed of light while the photon-baryon plasma moves roughly at the speed of sound, the net effect of modified perturbations of photons on the CMB power spectrum will be imprinted as a change in the phase shift as well as amplitude of Baryon Acoustic Oscillations (BAO). The process can be understood as follows:

In the $\Lambda \text{CDM}$ model, the phase shift and amplitude of BAO in the CMB power spectrum can be expressed as~\cite{Das:2020xke}:  
\begin{equation}
    \phi_\nu \approx 0.19\pi R_{\nu}, 1+ \Delta_\nu \approx 1 - 0.27 R_{\nu},
\end{equation}
where 
\begin{equation}
    R_{\nu} = \frac{\rho_{\nu}}{\rho_{\nu} + \rho_{\gamma}}
\end{equation}
 is the ratio of free-streaming neutrino energy density to the total radiation energy density. If we include the self-interaction between neutrinos, it would allow them to remain in thermal equilibrium with each other until relatively late times. As a result of this, the value of free-streaming neutrino fraction $R_{\nu}$ will get decreased relative to its $\Lambda$CDM value, depending on the total
number of neutrinos which are coupled at a particular time.  This would lead to a decrease in the phase shift and an increase in the amplitude of baryon acoustic oscillations. 
 The CMB multiple for a particular mode $k$ is given by~\cite{Das:2020xke}
\begin{equation}
\label{eq:l}
    l\approx \frac{(m\pi-\phi_\nu)}{\theta_\ast},~~{\rm with}~{\theta_\ast}=\frac{D^\ast_A}{r^\ast_s} \,,
\end{equation}
where $m\pi$ denotes the position of peaks, $\phi_\nu$ is the phase shift, $D^\ast_A$ is the distance to the surface of the last scattering from today, and $r^\ast_s$ is the radius of the sound horizon at the time of recombination. 
The $D^\ast_A$ and $r^\ast_s$ are expressed as a function of the Hubble parameter $H(z)$ as follows~\cite{Das:2020xke};\begin{eqnarray}
D^\ast_A = \int_{0}^{z^\ast}\frac{1}{H(z)} \,dz, \\ 
 r^\ast_s = \int_{z^\ast}^{\infty}\frac{c_s(z)}{H(z)} \, dz,
\end{eqnarray}
where $c_s(z) \approx 1/\sqrt{3}$ is the speed of sound in the baryon-photon plasma.
We can see from eq.~\ref{eq:l} that the decrease in  the phase shift $\phi_\nu$ due to self-interactions of neutrinos will shift the position of CMB multiple towards high $l$ values. In order to compensate for the shift to match with the observed power spectrum, we have to increase $\theta_\ast$. This can be achieved by increasing the value of $D^\ast_A$, while keeping $r^\ast_s$  unchanged.  In flat $\Lambda$CDM model, the Hubble constant evolves with redshift $z$ as $H(z) = H_0 \sqrt{\Omega_r (1+z)^4 + \Omega_m (1+z)^3 + \Omega_\Lambda}$, where $\Omega_m$, $\Omega_r$ and $\Omega_\Lambda$ corresponds to the fraction of the energy density acquired by matter, radiation and vacuum in the universe. If we slightly increase the value of $H_0$ such that there is increase in the value $H(z)$ at low red-shift while there is negligible change for $H(z)$ at high redshifts, we will be able to enhance $\theta_\ast$ such that the observed CMB multipole $l$ would remain unchanged. In this way, the presence of self-interacting neutrinos necessitates a higher value of $H_0$, thus alleviating the Hubble tension.

\begin{figure}[htbp]
\begin{center}
	\begin{tikzpicture}
	\begin{feynman}
		\vertex  (a) ;
		\vertex [left=of a] (b) {$\nu_i$};
		\vertex [right=of a] (c) {$\nu_i$};
		\vertex [below=of a] (d) ;
		\vertex [left=of d] (e) {$\nu_i$};
		\vertex [right=of d] (f) {$\nu_i$};
		\vertex [below=0.7 of a] (g) ;
		\vertex [right=.01 of g] (h) {$\phi$};
		
		\node[right =4 of g, fill=black, circle, inner sep=3pt] (i);  
		\vertex [above left=of i] (j) {$\nu_i$};
		\vertex [below left=of i] (k) {$\nu_i$};			
		\vertex [above right=of i] (l) {$\nu_i$};
		\vertex [below right=of i] (m) {$\nu_i$};

		\diagram*{
			(a) -- [] (b);
			(a) -- [] (c);
			(a) -- [scalar] (d);
			(d) -- [] (e);
			(d) -- [] (f);
			
			(i) --[] (j);	
			(i) --[] (k);
			(i) --[] (l);
			(i) --[] (m);		
				};	
	\end{feynman}
\end{tikzpicture}
  \caption{Feynman diagrams representing the non-standard interaction between neutrinos $\nu_i$ for $i= e,\mu,\tau$.}
  \end{center}
  \end{figure}
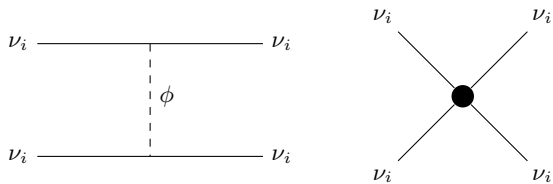
The  self neutrino interactions are governed by the following non-renormalizable interaction term:
\begin{equation}
    {\cal L} \supset  G^{ij}_{\rm eff} ({\bar \nu_i} \nu_i)({\bar \nu_i} \nu_i),
\end{equation}
where $G_{\rm eff}$ corresponds to effective coupling and $\nu_i = \nu_e, \nu_\nu,\nu_\tau$. In the early universe, this interaction can be mediated by heavy/light scalars as shown in Feynman diagram given in fig.~1. It has been found in ~{\cite{Cyr-Racine:2013jua,Lancaster:2017ksf} that the strength of self-interacting neutrino required to get the required value of Hubble constant can be categorized in two regimes, namely strongly-interacting neutrino (SI$\nu$) and moderately-interacting neutrino (MI$\nu$). The values of $G_{\mathit{eff}}$ in both regimes are given by~: 
\begin{equation}
G_{\mathit{eff}} = \begin{cases}
            (4.7^{+0.4}_{-0.6}{\rm~MeV})^{-2},~{\rm~SI\nu}  \\
             (89^{+171}_{-61}{\rm~MeV})^{-2},~{\rm~MI\nu}. 
              \end{cases}
\end{equation}
These values are subjected to severe constraints from different laboratory experiments as well as cosmological observations~\cite{Blinov:2019gcj,Lyu:2020lps}. However, we note that after taking into account all the constraints, there is a small amount of parameter space left for $\tau$-generation of neutrinos. Thus, it is interesting to explore whether the viable regime of $G_{\rm eff}$ is also consistent with the self-interaction strength required to explain the right value of KeV-sterile neutrino relic abundance via DW mechanism. 
\section{KeV-sterile neutrino dark matter}
In this section, we review the role of self-interacting neutrinos in generating the relic abundance of KeV-sterile neutrino DM. The KeV-scale sterile neutrino has been considered to be a popular warm DM candidate, alleviating all issues related to small scale structure of the universe. There exists numerous methods of producing sterile neutrinos in the early universe such as non-resonant Dodelson-Widrow mechanism~\cite{Dodelson:1993je}, resonant neutrino oscillations in the presence of lepton asymmetry~\cite{Shi:1998km}, inflaton decay~\cite{Shaposhnikov:2006xi}, decay of heavier particles~\cite{Roland:2014vba,Asaka:2006ek} etc. Given that the standard DW mechanism produces sterile neutrino DM without including a lot of ingredients from the early universe and physics beyond SM, it has been considered as one of the attractive mechanisms to generate the relic abundance of KeV-sterile neutrino. In the following subsections, we briefly discuss the calculation of relic abundance of sterile neutrino DM obtained through standard DW mechanism and modified DW mechanism in the presence of self-interacting neutrinos respectively.

\subsection{Standard Dodelson-Widrow Mechanism}
The standard  Dodelson-Widrow (DW) mechanism postulates the existence of an additional SM singlet neutrino  as realistic WDM candidate~\cite{Dodelson:1993je}. In the flavor basis, it can be written as a linear combination of active (SM) $\nu_i$ and sterile neutrino $\nu_4$, with physical eigenstate $\nu_s = \nu_i \sin \theta + \nu_4 \cos \theta$, with $\nu_i = \nu_e, \nu_\mu, \nu_\tau$. The angle $\theta$ measures the mixing between the SM and sterile neutrinos. For all practical purposes, we consider $\theta<<1$.

In the early universe, the active neutrinos remain in thermal equilibrium with all other particles while the sterile neutrinos do not any have any interaction with SM particles (except feeble interaction with active neutrinos). Therefore, it is assumed that the sterile neutrino has negligible initial abundance. As sterile neutrinos gets mixed with active neutrino at tree-level, the most efficient production method of sterile neutrino remains due to active to sterile  ($\nu_i \rightarrow \nu_s$) oscillations through a mechanism similar to the SM neutrino oscillations. Basically, while neutrino eigenstates propagate freely in the plasma for some time, they acquire a small component of sterile neutrino eigenstate. Eventually, the quantum mechanical “measurement” collapses the neutrino eigenstate into a sterile state with a small probability. This process continues until the active neutrinos decouple from the thermal plasma. After decoupling, the sterile neutrinos present at that time “freezes in” and left with a non-negligible relic abundance.

The production of KeV-sterile neutrino DM through DW mechanism can be described with the help of the Boltzmann equation in an expanding universe \cite{Kopp:2021jlk}
\begin{eqnarray}
&& \hskip -0.2in \left(\frac{\partial}{\partial t}-H E \frac{\partial}{\partial E}\right) f_{\nu_s}(E, t)=\left[\frac{1}{2} \sin ^{2}\left(2 \theta_{M}(E, t)\right) \Gamma(E, t)\right]\nonumber\\
&& \hskip 1.7in \times f_{a}(E, t), 
\end{eqnarray}
where $f_{a}(E, t)$ and $f_{\nu_s}(E, t)$ correspond to the time-dependent momentum distribution function of the active and sterile neutrino, respectively, $\theta_{M}(E, t)$ corresponds to the mixing angle in the matter, $\Gamma(E,t) = \frac{7 \pi}{24}G^2_F E T^4$ is the interaction rate of active neutrinos in the presence of SM weak interactions, and $H$ is the Hubble parameter.
The relic abundance is given by
\begin{equation}
 \Omega_{\nu_s}(0)= \frac{n_{\nu_s}(0)}{\rho_{\rm DM}} = \frac{m_{\nu_s} \rho_{\nu_s}(0)}{\rho_{\rm DM}(0)},
 \end{equation}
with
  $ n_{\nu_s}(0)  =  \int^{\infty}_{0} \frac{d^3E}{(2\pi)^3}f_{\nu_s} (E)$. 
With an appropriate choice of $m_{\nu_s}$ and the mixing angle $\theta_M$, the DW mechanism can produce enough sterile neutrinos to make up for the DM relic abundance observed today. However, this possibility has been ruled out by X-Ray observations and phase-space considerations.  The analysis of DM phase-space distribution in dwarf galaxies gives a lower bound on $m_{\rm DM}> 2$ KeV~\cite{Tremaine:1979we, Boyarsky:2008ju, Merle:2015vzu, Abazajian:2017tcc}.  The X-Ray observation has excluded almost the whole parameter space of $m_{\nu_s}-\sin^2{2\theta}$ required to explain  the relic abundance of sterile neutrino DM using DW mechanism \cite{Watson:2011dw, Horiuchi:2013noa, Perez:2016tcq, Dessert:2018qih, Ng:2019gch, calore2022constraints}. The resulting parameter space of $m_{\nu_s}-\sin^2{2\theta}$ is shown as the solid black line in fig~4. We can clearly see that the entire parameter space is ruled out by X-Ray observations. 
\subsection{Dodelson-Widrow Mechanism with Self-Interacting Neutrinos}
\label{sec:DWSI}
Recently, the modified Dodelson-Widrow mechanism has been proposed by considering self-interactions of active neutrinos mediated by scalar/vector particles~\cite{DeGouvea:2019wpf}. 
The interaction term of neutrinos with a new scalar mediator is given by
\begin{equation}
    {\cal L} \supset \frac{\lambda_\Phi}{2}\nu_i \nu_i \phi +h.c.
\end{equation}
 where $i=\nu_e,\nu_\mu,\nu_\tau$ corresponds to the generation of active neutrinos.

For active neutrino temperature $T$ and fixed neutrino energy $E=xT$, the distribution function of sterile neutrino as a function of temperature is given by
\begin{equation}
	\frac{df_{\nu_s}}{dz} =\frac{\Gamma \sin^2 2\theta_{eff}}{4 Hz} f_{\nu_i},\,
 \label{eq:dfdz}
\end{equation}
where
\begin{eqnarray}
	\sin^2 2\theta_{eff} \approxeq \frac{\Delta^2 \sin^2\theta}{\Delta^2 \sin^2 \theta +\frac{\Gamma^2}{4}+(\Delta\cos 2\theta - V_T)^2}.
\end{eqnarray}
Here  $z = \frac{\mu}{T}$ is a dimensionless variable with $\mu =1$ MeV, $\Delta = \frac{m^2_4}{2E}$ is the oscillation frequency of neutrinos in vacuum, $\Gamma$ corresponds to the total interaction rate for self-interacting active neutrinos, $\theta_{eff}$ is the effective mixing angle of sterile neutrino in the presence of self-interactions of active neutrino, $f_{\nu_s}$ and $f_{\nu_a}$ corresponds to the phase-space distribution function for sterile and active neutrinos respectively, and $V_T$ is the thermal potential. The active-neutrino self-interaction rate $\Gamma$ is given by~\cite{DeGouvea:2019wpf}:
\begin{equation}
\Gamma_\phi = \int \frac{d^3 p_{tar}}{(2\pi)^3} \frac{1}{e^{(E_{tar}/T)\,+\,1}}\, \sigma (\nu_a\nu_a \leftrightarrow \nu_a\nu_a)\, v_{\text{M\o{}ller}}.
\end{equation}
Here
$v_{\text{M\o{}ller}} = \sqrt{ {(\vec{v}_{\text{in}} - \vec{v}_{\text{tar}})}^2 - {(\vec{v}_{\text{in}} \times \vec{v}_{\text{tar}})}^2 } $  
is the M\o{}ller velocity between the incoming and the target particle
and $\sigma$ is the cross-section given by:
\begin{equation}
\sigma (\nu_a\nu_a \leftrightarrow \nu_a\nu_a) = 
 \frac{\lambda^4_\phi\,s}{32\pi \left( {(s-m_\phi^2)}^2+m^2_\phi\gamma^2_\phi  \right)}.
\end{equation}
The value of $s$ can be calculated by using $s=2E_{\text{in}}\,E_{\text{tar}}(1-\cos \theta)$, where $E$ is the energy of scattering neutrinos and $\theta$ is the scattering angle.

In the limiting case, it would follow:
\begin{equation}
\Gamma_\phi = \begin{cases}
              \frac{7 \pi \lambda^4 E_1 T^4}{864 m^4_{\phi}}, &\text{if $m_\phi>>T$ }   \\
              \frac{\lambda^2 m_\phi^2 T}{8\pi E_1^2} \left(\ln(1+e ^y)-y\right), & \text{if $m_\phi\lesssim T $}
              \end{cases}
\end{equation}
where $y = \frac{m_\phi^2}{4E_1 T}$. The total interaction rate will be given by $\Gamma= \Gamma_{\rm SM}+ \Gamma_\phi+\Gamma^c_\phi$, where $\Gamma_{SM} \approx G^2_F E T^4$, with $G_f$ being the Fermi constant.

Similarly, the total thermal potential $V_T$ will have contribution from the standard model weak interactions as well as new self-interaction among neutrinos. The standard model thermal potential is given by 
\begin{eqnarray}
&& {\hskip -0.25in} V^{SM}_T \approx \frac{ G_F E T^4}{M^2_W}.
\end{eqnarray}
For a mediator mass $m_\phi$, the thermal potential arising from self-interaction of neutrinos is given by
\begin{eqnarray}
  && \hskip -0.3in  V_{T}^{\phi}(E, T) =\frac{\lambda_{\phi}^{2}}{16 \pi^{2} E^{2}} \int_{0}^{\infty} d p\Biggl[\left(\frac{m_{\phi}^{2} p}{2 \omega} L_{2}^{+}(E, p)-\frac{4 E p^{2}}{\omega}\right) \nonumber\\
    && \frac{1}{e^{\omega / T}-1} +\left(\frac{m_{\phi}^{2}}{2} L_{1}^{+}(E, p)-4 E p\right) \frac{1}{e^{p / T}+1}\Biggr],
\end{eqnarray}
where
\begin{eqnarray}
&& 	\hskip -0.3in L_{1}^{+}(E, p) =\ln \frac{4 p E+m_{\phi}^{2}}{4 p E-m_{\phi}^{2}},\nonumber\\
&& \hskip -0.3in	L_{2}^{+}(E, p) =\ln \frac{\left(2 p E+2 E \omega+m_{\phi}^{2}\right)\left(2 p E-2 E \omega+m_{\phi}^{2}\right)}{\left(-2 p E+2 E \omega+m_{\phi}^{2}\right)\left(-2 p E-2 E \omega+m_{\phi}^{2}\right)}, \nonumber\\ && {\rm with}~w =  \sqrt{p^2+m_\phi^2}.
\end{eqnarray}
In the low/high temperature limit, it  takes the following form:
 \begin{equation}
 	V_{T}^{\phi}(E, T)=
 	                  \begin{cases}
 	                 \frac{-7 \pi^2 \lambda^2_\phi E T^4}{90 m^4_\phi}, & \text{if $m_\phi>> T$}\\
 	              	 \frac{\lambda^2_\phi T^2}{16 E}, & \text{if $m_\phi << T$}.
 	                 \end{cases}
\end{equation}
As most of the production of KeV-sterile neutrino occurs in the temperature range from T=10 GeV to T = 0.1 MeV (BBN limit), the distribution function will given by integrating eq.~(\ref{eq:dfdz}) in the desired temperature range. This follows~\cite{Chichiri:2021wvw}:
\begin{equation}
f_{\nu_s} (E) = \int^{z=10}_{z=0.0001} \frac{d f_{\nu_s}}{dz} dz,
\end{equation}
   with  $z={\rm~MeV}/T$. The final present day number density of sterile neutrino dark matter will be given by integrating $f_{\nu_s}(E)$ over the entire energy range as follows:
   \begin{equation}
   n_{\nu_s}(0)  =  \int^{\infty}_{0} \frac{d^3E}{(2\pi)^3}f_{\nu_s} (E).
   \end{equation}
   Rewriting in terms of $x=E/T$, we get
      \begin{equation}
   n_{\nu_s}(0) = \frac{T^3}{\pi^2}\int^{\infty}_{0} {dx}f_{\nu_s} (x).
   \end{equation}
By expressing number density of sterile neutrino in terms of active neutrino, with $\nu_i(0) =112~{\rm cm}^3$ per active neutrino, we get~\cite{Chichiri:2021wvw}:
 \begin{equation}
   n_{\nu_s}(0) = n_{\nu_i}(0) \left(\frac{g_*}{10.75}\right)^{-1} \frac{2}{3\zeta(3)}\int^{\infty}_{0} {dx}f_{\nu_s} (x).
   \end{equation}
The fraction of sterile neutrino DM relic abundance will be given by~\cite{Chichiri:2021wvw}:
 \begin{equation}
 \Omega_{\nu_s}(0)= \frac{n_{\nu_s}(0)}{\rho_{\rm DM}} = \frac{m_{\nu_s} \rho_{\nu_s}(0)}{\rho_{\rm DM}(0)},
 \end{equation}
 where $m_{\nu_s}$ is the mass of sterile neutrino, ${\rho_{\rm DM}} = 0.26 {\rho_{c}}$, with ${\rho_{c}}= 1.05 \times 10^{-5} h^2$ being the critical density of the universe with $h =0.7$.

In the next section, we use this calculation to obtain the relic abundance of KeV-sterile neutrino for the preferred range of $G_{\rm eff}$. 
\section{Synergy between self-interacting neutrino and KeV-Sterile neutrino DM}
 In this section,  we have explored the synergy between the parameter space of $G^{\tau \tau}_{\rm eff}$ ($g^{\tau \tau}_{\phi}$ vs $m_{\phi}$) required to assist in the successful production of keV-sterile neutrino DM and the one required to alleviate $H_0$ tension.

 As we mentioned in section II, the strength of neutrino self-interaction required to get the desired value of $H_0$ from CMB observations prefer the value of  $G_{\rm eff}$ to be $(4.7^{+0.4}_{-0.6}{\rm~MeV})^{-2}$, dubbed as ``strongly" interacting region, and $(89^{+171}_{-61}{\rm~MeV})^{-2}$ dubbed as ``moderately” interacting neutrinos regime~\cite{Lancaster:2017ksf}. It has been demonstrated in \cite{Blinov:2019gcj} that for these fixed values of $G_{\rm eff}$, most of the parameter space of  $g_{\rm \phi/A_\nu}$ (neutrino interaction coupling to scalar/vector mediators) vs $m_{\rm \phi/A_\mu}$ (mass of scalar/vector mediator) is completely ruled out by severe laboratory as well as cosmological constraints~\cite{Blinov:2019gcj,Lyu:2020lps}. In case of neutrino self-interactions mediated through vector fields $A_\mu$, the entire  parameter space of $g_{A_\nu}-m_{A_\nu}$ corresponding to $G^{MI\nu}_{\rm eff}$ and  $G^{SI\nu}_{\rm eff}$ is ruled out by constraints from Big-Bang Nucleosynthesis (BBN)~\cite{Blinov:2019gcj} while  in the case of scalar mediators, the laboratory constraints are specific to a particular flavour of neutrino. The parameter space of $g^e_{\phi}$ and $g^\mu_{\phi}$ is disfavoured by constraints from laboratory constraints as well BBN observations. However, the substantial amount of $g_{\phi}-m_{\phi}$ parameter space is left for $\tau$-generation of neutrinos  even after taking into account all the constraints from astrophysical/cosmological considerations and colliders~\cite{Blinov:2019gcj}. Thus we will only explore the allowed parameter space of $g^{\tau \tau}_{\phi}$ vs $m_{\phi}$. 
 



\begin{figure}[t]
     \centering
\hskip -0.2in \includegraphics[width=.5\textwidth, height =0.5\textwidth]{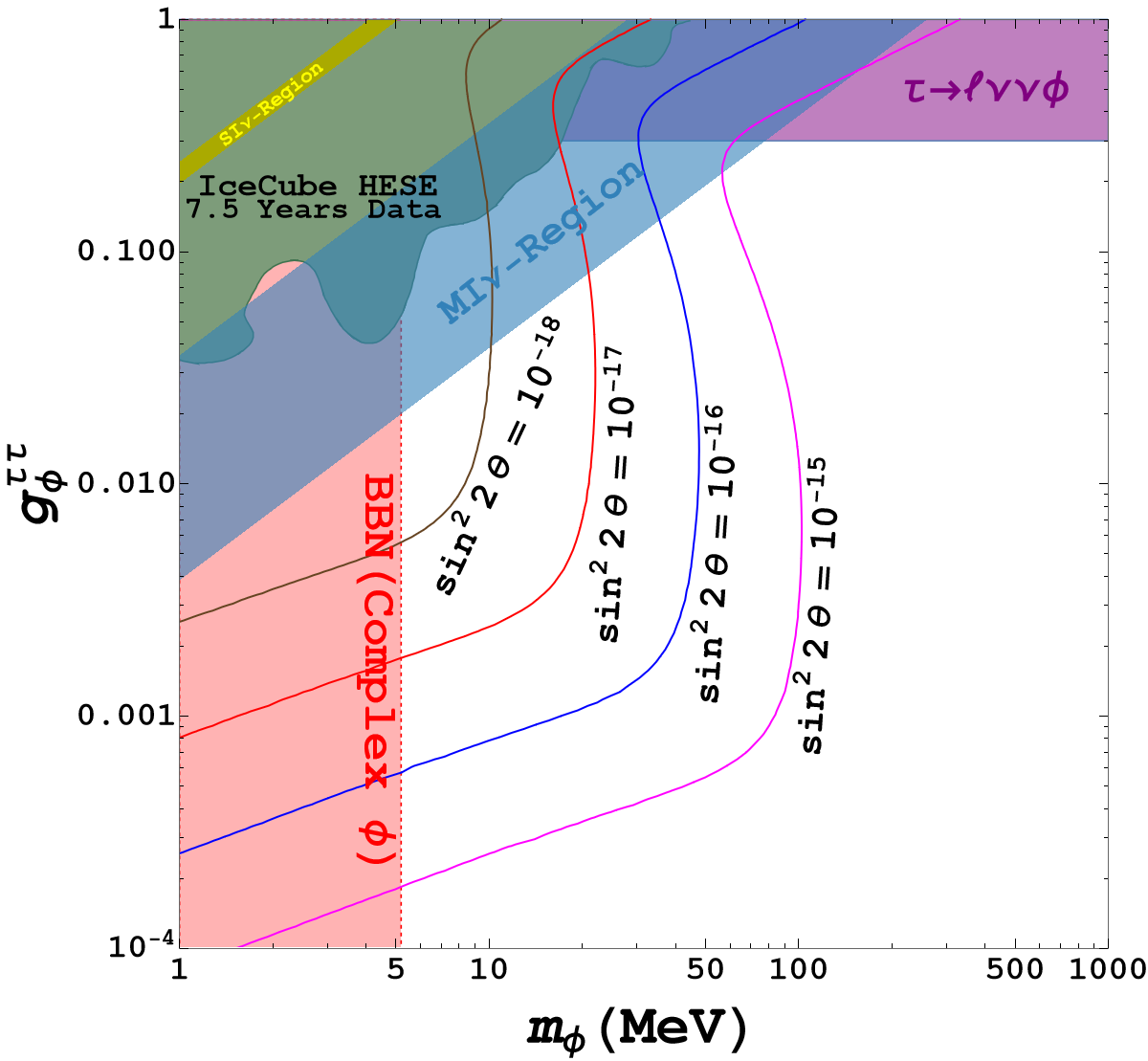}
     \caption{The four differently coloured curves correspond to the relic abundance of sterile neutrino $\Omega_{\nu_s}\sim 0.12$ for (i) $\sin^2{2\theta} = 10^{-18}$, $m_{\nu_s}= 10$ KeV, (ii) $\sin^2{2\theta} = 10^{-17}$, $m_{\nu_s}= 10$ KeV, (iii)  $\sin^2{2\theta} = 10^{-16}$, $m_{\nu_s}= 10$ KeV, and (iv)  $\sin^2{2\theta} = 10^{-15}$, $m_{\nu_s}= 10$ KeV. The orange shaded represents the region ruled out by constraints from BBN~\cite{Blinov:2019gcj}. The purple shaded region shows the parameter space excluded from bounds on the decay rate of $\tau \rightarrow l \nu \nu \phi$~\cite{Lessa:2007up}. The green shaded region shows the excluded parameter space from astrophysical flux of high energy neutrino obtained by using 7.5 years of IceCube data~\cite{Esteban:2021tub}. The dark green and blue shaded bands correspond to preferred range of $G_{\rm eff}$ in the ``strongly" interacting region, and ``moderately” interacting neutrinos regime respectively~\cite{Lancaster:2017ksf,Blinov:2019gcj}.}
     \label{fig:my_label}
\end{figure}

\begin{figure}
     \centering
\hskip -0.2in \includegraphics[width=.5\textwidth, height =0.5\textwidth]{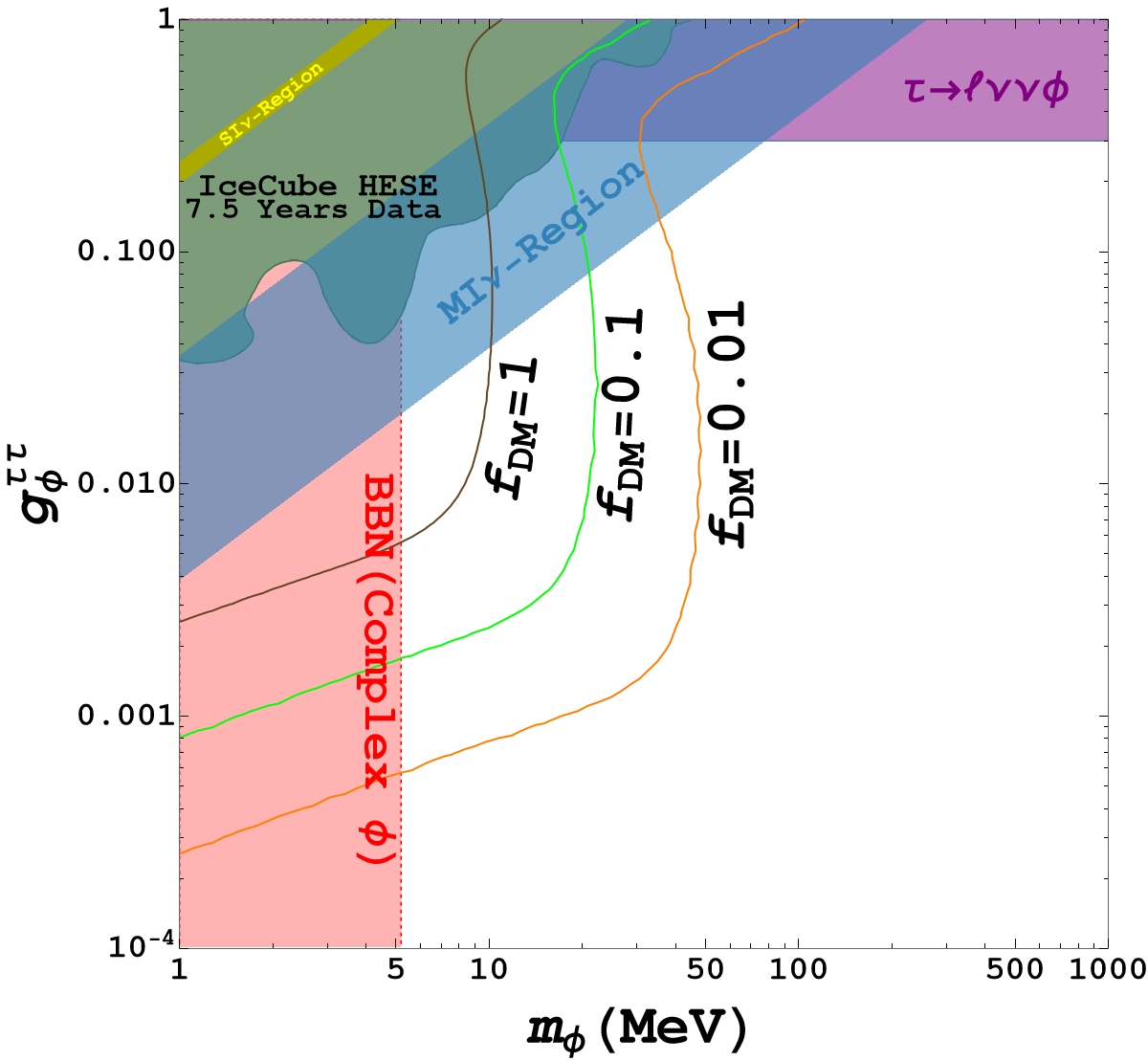}
     \caption{The three differently colored curves correspond to the resulting parameter space of $g^{\tau \tau}_{\Phi}$ vs $m_{\phi}$ for the smallest allowed value of $ \sin^2{2\theta} = 10^{-18}$ and $m_{\nu_s} = 10$ KeV by assuming the fraction of sterile neutrino DM to be (i)$ f_{\nu_s} = \Omega_{\rm DM}$, (ii) $f_{\nu_s} = 0.1 \Omega_{\rm DM}$, and (iii) $ f_{\nu_s} = 0.01 \Omega_{\rm DM}$. All the other shaded regions are same as in fig.~2.}
     \label{fig:my_label}
\end{figure}
The SI$\nu$ and MI$\nu$ region of the parameter space has been shown as the green and blue shaded region in fig.~2. In addition to that, there are severe constraints from laboratory and cosmological considerations.  This has  been depicted clearly in fig.~2. The scalar mediator gets in thermal equilibrium with neutrinos through the process $\phi\rightarrow \nu_i \nu_i$ before the neutrino decoupling temperature $T_{\rm dec}$. If it remains relativistic throughout the period between $T_{\rm decay}$ and BBN, it will lead to $\Delta N_{\rm eff}\geq 1$ for complex $\Phi$. Thus, the massive scalar mediator needs to get the Boltzmann suppression before the onset of BBN. These constraints have been calculated in~\cite{Blinov:2019gcj} and give a  bound on $m_{\Phi} \geq 5.2$ MeV for complex scalar mediator. The ruled-out region is shown in the orange shaded band in fig.~2. The constraints from laboratory comes from the  decay channel of $\tau$-lepton to light scalars as given by $\tau \rightarrow l\nu\nu \phi$. The experimental bounds on $\tau$ decay rate puts a bound on the couplings $|g^{\tau \tau}_{\phi}|^2 < 5.5 \times 10^{-2}$~\cite{Lessa:2007up}, which gives $g^{\tau \tau}_{\phi} \leq 0.3$ for light scalar mediators. The ruled out parameter space from this bound is shown as purple shaded band in fig.~2. Further, the self-interaction of tau-neutrinos can also be probed from the detection of high-energy neutrinos by IceCube collaboration~\cite{Esteban:2021tub}. The scattering of high-energy astrophysical neutrinos with CMB neutrinos passing through the Earth redistribute their energies, resulting in dips/bumps in the observed spectrum of the diffuse astrophysical neutrino flux background. The astrophysical flux of high-energy neutrinos obtained by using 7.5 years of IceCube data~\cite{Esteban:2021tub} excludes the green shaded region of $g^{\tau \tau}_{\phi}-m_{\rm \phi}$ parameter space shown in fig.~2. We can also see in fig.~2 that the SI$\nu$ region of the parameter space is completed ruled out by the aforementioned constraints. 

The relic abundance of KeV-sterile neutrino for the particular value of $G^{{\rm MI}\nu}_{\rm eff}$ has been calculated by using modified DW mechanism explained in section III.B. By considering the requirement that the sterile neutrino accounts for the entire dark matter of the universe, we sketch out the resulting parameter space of $g^{\tau \tau}_{\phi}$ vs $m_{\phi}$ for the fixed mass of the sterile neutrino $m_{\nu_s}\sim 10$ KeV and different  values of $\sin^2{2\theta}$ in fig.~2. The four differently colored curves correspond to the relic abundance of sterile neutrino $\Omega_{\nu_s}\sim 0.12$ for (i) $\sin^2{2\theta} = 10^{-18}$, $m_{\nu_s}= 10$ KeV, (ii) $\sin^2{2\theta} = 10^{-17}$, $m_{\nu_s}= 10$ KeV, (iii)  $\sin^2{2\theta} = 10^{-16}$, $m_{\nu_s}= 10$ KeV, and (iv)  $\sin^2{2\theta} = 10^{-15}$, $m_{\nu_s}= 10$ KeV. We can see that there exists points in the parameter space of MI$\nu$ region which also give the right value of relic abundance for the mass of sterile neutrino to be 10 KeV and $\sin^2{2\theta}$ ranging between $10^{-15}-10^{-18}$. In different words, we have shown that getting the right value of relic abundance for 10 KeV sterile neutrino would constraint the mixing angle $\sin^2{2\theta} \leq 10^{-15}$. 

\begin{figure*}[htbp]
\includegraphics[width=.85\textwidth]{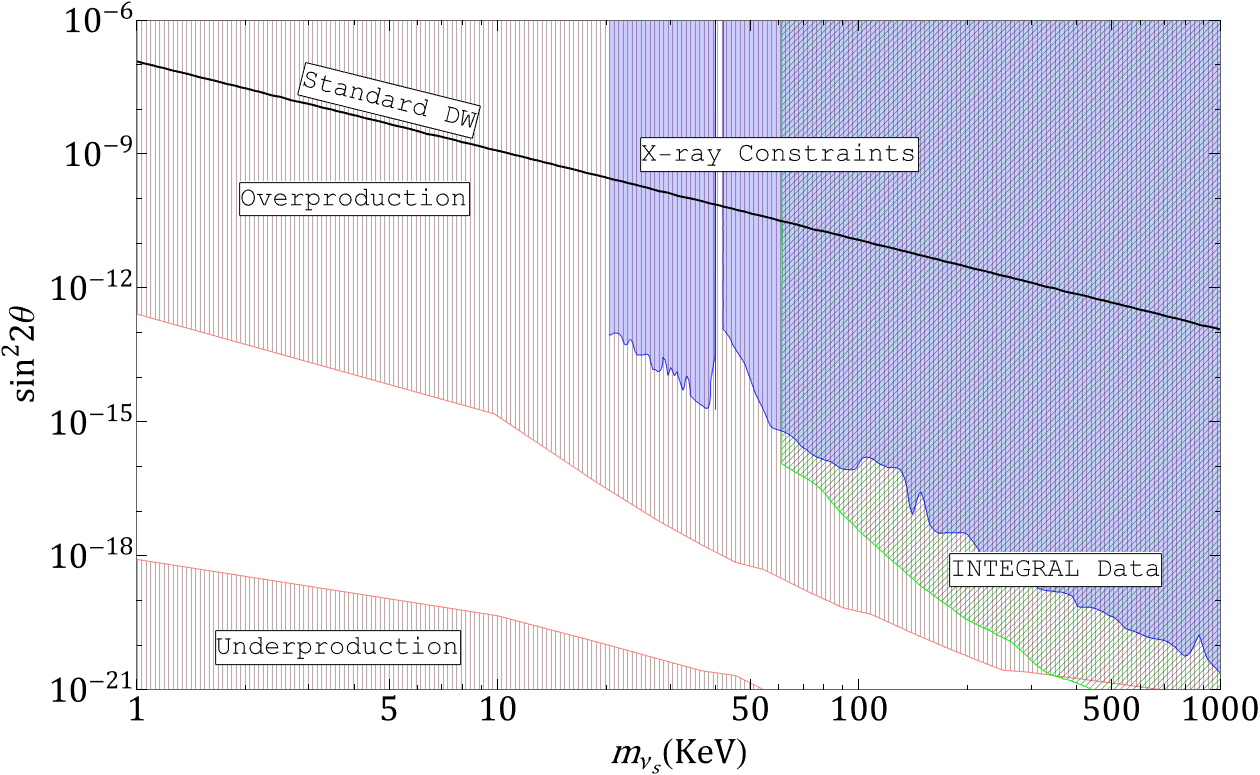}
     \caption{ The vertically spaced pink coloured lines correspond to the regions for which the relic abundance of sterile neutrino becomes under-abundant ($\Omega_{\nu_s}\leq \Omega_{\rm DM}$) and over-abundant ($\Omega_{\nu_s}\geq \Omega_{\rm DM}$) respectively. The region in between pink lines would constitute the entire DM of the universe. The blue shaded vertically spaced lines correspond to region excluded by X-Ray constraints from NuSTAR observations~\cite{roach2022long,Boyarsky:2007ge}. The green shaded region corresponds to the region excluded by X-Ray constraints obtained from the analysis of 16 years INTEGRAL data~\cite{Calore:2022pks}. We can also see that the entire available region is safe from X-Ray constraints.}
     \label{l3}
\end{figure*}

\begin{figure*}[t]
     \centering
 \includegraphics[width=.85\textwidth]{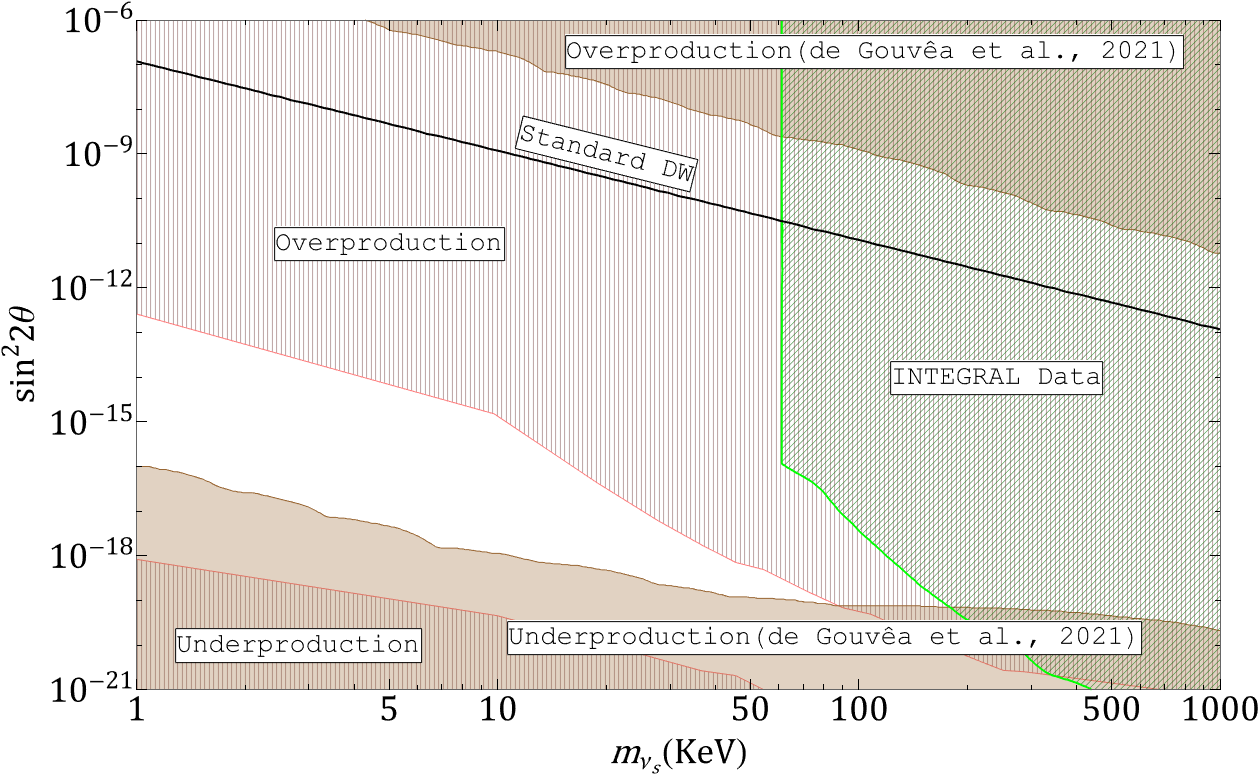}
     \caption{For direct comparison, we have shown the over-abundant and under-abundant region of $m_{\nu_s}-\sin^2{2\theta}$ parameter space obtained by \cite{DeGouvea:2019wpf} as brown shaded region. Our results are shown as pink shaded regions as in fig.~4. The green shaded region corresponds to X-Ray constraints from 16 years of INTEGRAL data~\cite{Calore:2022pks}. We can clearly see that both over-abundant and under-abundant region of sterile neutrino DM in our case get shifted towards more suppressed value of mixing angle. As a result of this, our entire available parameter space becomes almost safe from X-Ray constraints obtained using INTEGRAL data~\cite{Calore:2022pks}.}
     \label{l2}
\end{figure*}

We have also explored availability of the parameter space by assuming that the KeV-sterile neutrino might contribute to a fraction of the overall DM present in the universe. Thus, we have plotted the resulting parameter space of $g^{\tau\tau}_{\phi}$ vs $m_{\phi}$ for smallest allowed value of $ \sin^2{2\theta} = 10^{-18}$ and $m_{\nu_s} = 10$ KeV by assuming the fraction of sterile neutrino DM to be (i)$ f_{\nu_s} = \Omega_{\rm DM}$, (ii) $f_{\nu_s} = 0.1 \Omega_{\rm DM}$, and (iii) $ f_{\nu_s} = 0.01 \Omega_{\rm DM}$ in fig.~3. Our results indicate that allowed parameter space of $g^{\tau \tau}_{\phi} - m_{\phi}$ starts diminishing by choosing a small fraction of the DM. Thus, if we need the sterile neutrinos to contribute a smaller fraction of the DM, we would need even more suppressed mixing angle. 

Finally, we have discretized and scanned over the neutrino mass and mixing in the range ${\rm KeV}<m_{\nu_s}<{\rm~MeV}$ and $10^{-21}<\sin^2{2\theta}<10^{-6}$ suited to obtain $\Omega_{\rm \nu_s} = \Omega_{\rm DM}$ for all the values of $g^{\tau \tau}_{\phi}$ and $m_{\phi}$ which satisfy $28~{\rm  MeV}^{-2}\leq G_{\rm eff} \leq 260$ MeV$^{-2}$ (except  $m_{\Phi}\leq 5.2$ MeV  and $g^
{\tau\tau}_{\phi}\geq 0.3$ as excluded by BBN and $\tau$-decay constraints respectively).

For the two-body final state decay mode in which sterile neutrino DM decays $\nu_s \rightarrow \nu_i \gamma$, where $\nu_i$ is a lighter and an active neutrino with $m_{\nu_s} \gg m_{\nu_i}$, the emitted phone carries energy equals to half of the sterile neutrino rest energy. As the emitted energy of photon lies in KeV-MeV range, this can be observed by various X-Ray observations.  The constrains are obtained on the value of the mixing angle $\sin^2{2\theta}$ for the sterile neutrino mass range 4 KeV $< m_{\nu_s} <$ 40 KeV by analyzing the data from Nuclear Spectroscopic Telescope Array (NuSTAR~\cite{roach2022long}), and for the sterile neutrino mass range 40 KeV $< m_{\nu_s} <$ 14 MeV by analyzing INTEGRAL/SPI data~\cite{Boyarsky:2007ge}. These constraints are shown as blue shaded region in fig.~4.  In a recent study \cite{Calore:2022pks}, the authors analyzed 16 years of X-Ray data from the Soft Photon Imager(SPI), the  high-resolution gamma-ray spectrometer on board the International Gamma-Ray Astrophysics Laboratory (INTEGRAL), to constrain the $m_{\nu_s}-\sin^2{2\theta}$ parameter space. Their constraints are much stronger then the constraints obtained in \cite{Boyarsky:2007ge} for heavy value of $m_{\nu_s}$. The INTEGRAL constraints are shown as green shaded region in fig.~4 and fig.~5. 
Our results are shown in fig.~4. We have hatched in the region for which the relic abundance becomes under-abundant ($\Omega_{\nu_s}\leq \Omega_{\rm DM}$) and over-abundant ($\Omega_{\nu_s}\geq \Omega_{\rm DM}$) by  pink colored vertically spaced lines in fig.~4. The region in between two pink lines would constitute the entire DM of the universe. This intermediate region corresponds to the mixing angle $\sin^2{2\theta}\sim 10^{-13}$-$10^{-21}$ for mass of sterile neutrino $m_{\nu_s}\sim 1-1000$ KeV. We also note that the entire available region becomes safe from X-Ray constraints. Thus, we can conclude that the requirement of KeV-sterile neutrino DM for a specific range of $G_{\rm eff}$ would constrain the mixing angle $\sin^2{2\theta}$ between  $10^{-13}$-$10^{-21}$.

In fig.~5, we compare our results with $m_{\nu_s}-\sin^2{2\theta}$ parameter space obtained by \cite{DeGouvea:2019wpf} by considering $10^{-6}\leq g^{\mu \mu}_{\phi} \leq 1$ and $m_{\phi}\sim 1-10^4$ MeV for $\mu$-neutrino case (except the region excluded by Kaon decay constraints and BBN observations). As the Kaon decay constraints mostly exclude the range of couplings $g^{\mu \mu}_{\phi}\geq 0.01$ for $m_{\phi} \sim 1-100$ MeV, we should note that this is almost the same range we have used to obtain the value of $G^{{\rm MI}\nu}_{\rm eff}$. Thus our results are complementary to the results obtained in \cite{DeGouvea:2019wpf}. For direct comparison, we have shown the overabundant and underabundant region of $m_{\nu_s}-\sin^2{2\theta}$ parameter space obtained by \cite{DeGouvea:2019wpf} as brown shaded region in fig.~5. 
It can be clearly seen that both overabundant and underabundant region of sterile neutrino DM in our work gets shifted towards more suppressed value of mixing angle. As a result of this, our entire available parameter space is safe from X-Ray constraints obtained using 16 years of INTEGRAL data~\cite{Calore:2022pks}.

Thus, we conclude that we are able to obtain the parameter space of $m_{\nu_s}-\sin^2{2\theta}$ which is consistent with the preferred range of $G_{\rm eff}$ and the recent results from X-Ray observations.

\section{Consistent phenomenological model}
 In this section, we present a phenomenological model which can give the required mass of SM neutrino and KeV-scale mass of the right-handed (sterile) neutrino DM along with tiny mixing angle as considered in this work. In standard scenario such as type-I see-saw mechanism, the required mass of SM neutrino is obtained by considering additional heavy Majorana right-handed neutrino with mass around $10^9$ GeV~\cite{Sarkar:2008xir}. However, in our work, we need to consider KeV-scale mass of right-handed neutrino and a very tiny Yukawa coupling so that the mixing angle between SM neutrino and right-handed neutrino turns out to be very small. Thus, the standard scenarios would not be able to produce the required mass of the SM neutrino.  Additionally, the Dirac nature of SM (left-handed) neutrino has  been disfavoured from the requirement of getting $\Delta N_{\rm eff}\le1$~\cite{Blinov:2019gcj}. Hence, one needs to  consider the Majorana nature of the SM neutrino.
Overall, it seems challenging to obtain this kind of spectrum of masses and coupling in typical models involving right-handed neutrinos. 
In this section, we have tried to explain this spectrum in the context of a very general $N=1$ supergravity framework with supersymmetry breaking scale around O(10) TeV.  Below, we have sketched out a toy supergravity model which would naturally explain the existence of light SM neutrino along with KeV scale right-handed neutrino and tiny mixing angle.\footnote{ We have not worked out the full details of the model. We have only mentioned various terms relevant to neutrinos. A specific class of detailed supergravity model is presented in~\cite{Dhuria:2011ye,Dhuria:2012bc,Dhuria:2013syh}.}
  
 In supergravity models, all the interaction terms are obtained from the renormalizable as well as non-renormalizable operators present in the K\"{a}hler potential and superpotential. 
The model generally also includes a set of hidden-sector superfields which are singlets under the gauge group of the SM and play an important role in breaking supersymmetry.  The terms involving the fermion mass and Yukawa interactions of right-handed neutrinos are given in the superpotential:
 \begin{equation}
 W \supset y_N {\hat N} {\hat L} {\hat H_u} + m_N {\hat N} {\hat N}.
 \label{ymstd}
 \end{equation}
The mass of supersymmetric counterparts such as right-handed sneutrino is determined by non-renormalizable operators involving the interaction between visible and hidden-sector superfields in the K\"{a}hler potential
 \begin{equation}
{\cal K}  \supset \frac{{\hat X^\dagger_i} {\hat X_i}} {M_p} \left({\hat N}^{\dagger} {\hat N}\right).
 \label{kstd}
 \end{equation}
Here, (${\hat N}, {\hat L}, {\hat H_u} $) and ${\hat X_i}$ correspond to the visible sector matter superfields and hidden-sector superfields respectively. We can obtain the mass of supersymmetric scalar particle by integrating out F-term of the hidden-sector superfield.

From eqs.~(\ref{ymstd}) and (\ref{kstd}), we can see that the natural value of $y_N$ would be ${\cal O}(1)$ and the natural value of $m_N$ would be either zero or of the order of Planck scale. This is similar to $\mu$-problem in supersymmetric theories and it has been addressed by considering the famous Giudice-Masiero mechanism~\cite{Giudice:1988yz}. This mechanism is based on the idea that both hidden and visible sector superfields transform non-trivially under a new global symmetry~G. This symmetry would forbid the $\mu$-term in the superpotential, while allowing the same from  higher dimensional operators in the K\"{a}hler potential. The similar mechanism has also been adopted to generate the small value of neutrino mass from the non-renormalizable operator in the K\"{a}hler potential~$({\cal K} \supset {{\hat X_i}^\dagger  {\hat N} {\hat N}}/{M_P})$~\cite{Arkani-Hamed:2000oup}. 


In this work, we consider a Giudice-Masiero like mechanism to obtain the KeV scale right-handed neutrino and suppressed Yukawa coupling strength. We assume that charges of hidden and visible sector superfields under global symmetry $G$ are chosen in a way such that both the right-handed neutrino mass and Yukawa interaction term can be obtained from the following non-renormalizable operators in the K\"{a}hler potential:
\begin{eqnarray}
&&  {\cal K} \supset \frac{{\hat X_i}  {\hat N}^{\dagger} {\hat N}} {M_p} \left(1+ \frac{{\hat X^\dagger_i} {\hat X_i}}{M_p}\right) + \frac{{\hat X_i} {\hat X_i} {\hat l}^{\dagger} {\hat l}} {M^2_p} \left(1+ \frac{{\hat X^\dagger_i} {\hat X_i}}{M_p}\right) \nonumber\\
 && +  \frac{{\hat X^\dagger_i}  {\hat N} {\hat L} {\hat H_u}}{M^2_p} \left(1+ \frac{{\hat X^\dagger
 _i} {\hat X_i}}{M_p}\right) +h.c.
 \label{eq:kterms}
\end{eqnarray}
The hidden-sector superfield can be expanded  as ${\hat X_i} = X_i + \theta{\tilde \psi_i}  + \theta^2 F_{X_i} $, with $X_i$ being the scalar component of the superfield, ${\psi_i}$ being the fermion component of the right handed neutrino, and $F_{X_i}$ being the F-term of ${\hat X_i}$. Similarly, we can expand the neutrino superfields as: ${\hat N} = N + \theta{\tilde N}  + \theta^2 F_{N} $, ${\hat L} = l + \theta{\tilde l}  + \theta^2 F_{l}$, with $N$ being the singlet right-handed neutrino, $l^T = (\nu_e, e)^T$ being the lepton doublet, ${\tilde N},{\tilde l}$ being the scalar component and $F_{N}, F_{l}$ being the F-term of right-handed and left-handed neutrino superfield respectively.  The Higgs superfield will be expanded as:  ${\hat H_u} = H_u + \theta{\tilde H_u}  + \theta^2 F_{H_u}$ with  $H_u$ and ${\tilde H_u}$ being the Higgs scalar field and higssino fermion field respectively. The terms in the Lagrangian are obtained from the  K\"{a}hler potential by using $\int d^2\theta d^2{\bar\theta}{\theta}^2 {\bar\theta}^2 = 1$.

As we can see in eq.~(\ref{eq:kterms}), the non-renormalizable term $\frac{{\hat X_i}  {\hat N}^{\dagger} {\hat N}} {M_p}$ would not generate the mass term for Majorana neutrino as it would give $\int d^2\theta d^2{\bar\theta}{\theta}^4 = 0$. Thus, the Majorana mass of right-handed neutrino can be calculated from the next-order non-renormalizable term by giving VEV to $\langle X_i\rangle$ and F-term component of ${\hat X_i}$:
 \begin{equation}
 \int d^2\theta  \theta^2 {\bar \theta}^2 \left(\frac{F^*_{X_i} \langle X_i \rangle^2}{M^3_p}\right) N N \longrightarrow m^M_{\nu_s} =  \frac{F^*_{X_i}\langle X_i \rangle^2}{M^3_p}.
 \end{equation}
 Similarly, the Majorana mass of left-handed neutrino will be given as
  \begin{equation}
 \int d^2\theta  \theta^2 {\bar \theta}^2 \left(\frac{F^*_{X_i} \langle X_i\rangle^3}{M^4_p}\right) l l  \longrightarrow  m^{M}_{\nu_l} =\frac{F^*_{X_i}\langle X_i \rangle^3}{M^4_p}.
\end{equation}
The Yukawa interaction term can be calculated from the first order non-renormalizable term  $\frac{{\hat X^\dagger_i}  {\hat N} {\hat L} {\hat H_u}}{M^2_p}$. It will be given as
\begin{equation}
 \int d^2\theta  \theta^2 {\bar \theta}^2 \left(\frac{F^*_{X_i}}{M^2_p}\right) N l H_u  \longrightarrow y_N =  \frac{F^*_{X_i}}{M^2_p}.
 \end{equation}
With this, the Dirac Neutrino mass will be given by:
 \begin{equation}
m^{D}_{\nu_l} = y_N v_u,
 \end{equation}
 and the value of the mixing angle will be given by
\begin{equation}
\tan 2\theta = \frac{y_N v_u}{m_{\nu_s}},
\end{equation}
where $v_u = v \sin\beta$ with $v=246$ GeV and $\sin\beta \sim {\cal O}(1)$. Given that $y_N <<1$, we can get $\tan 2\theta \approx \sin 2\theta$. This gives $\sin^22\theta \sim \frac{y^2_N v^2_u}{m^2_{\nu_s}}$.
 
In the above results, we can see that the scale of masses of neutrinos and Yukawa coupling depends on the scalar VEV ($\langle X_i$) and F-term VEV ($F_{X_i}$) of the hidden superfield. In gravity mediated supersymmetric models, the supersymmetry breaking scale is generally given by $F_{X_i}/M_p$~\cite{Martin:1997ns}. Given that we have not seen any signatures of supersymmetry at TeV scale, we push and keep the supersymmetry breaking scale to be ${\cal O}(10)$ TeV. This gives $F_{X_i} = 10^{22}$ GeV. The scalar VEV of hidden supersymmetric field is generally chosen to be near Planck scale, we consider $\langle X_i\rangle = 10^{-4} M_p$. By considering these values, we calculate the values of neutrino mass and couplings. The spectrum is given in Table~1. If we do the diagonalization of neutrino mass matrix involving $\nu_s$ and $\nu_l$, the eigenvalues will be dominated by $m^{M}_{\nu_s}$ and $m^{M}_{\nu_l}$. Thus, we consider the Majorana nature of both the right-handed (sterile) and left-handed (SM) neutrino.

\begin{table}[htbp]
    \centering
    \begin{tabular}{ |l  | l|l | } 
 \hline
Particle & Mass & Scale \\ 
  \hline
Majorana right-handed (sterile) neutrino & $\frac{F^*_{X_i}\langle X_i \rangle^2}{M^3_p}$ & KeV \\
& &\\ 
  \hline
Majorana left-handed (SM) neutrino & $\frac{F^*_{X_i}\langle X_i \rangle^3}{M^4_p}$ & 0.1 eV \\  
& &\\ 
  \hline
 Yukawa coupling & $\frac{F^*_{X_i}}{M^2_p}$ & $10^{-15}$ \\  
 & &\\ 
  \hline
 Dirac neutrino & $\frac{F^*_{X_i}}{M^2_p} v_u $ & $10^{-4}$ eV \\  
 & &\\ 
 \hline
  $\sin^2{2\theta}$ &  $\frac{y^2_N v^2_u}{m^2_{\nu_s}}$  & $10^{-14}$ \\  
  \hline
\end{tabular}
\caption{Mass/coupling parameters and their values.}
\label{table:1}
\end{table}

Finally, we have shown that by choosing $O(10)$ TeV supersymmetry breaking scale (and corresponding $F_{X_i}$), we can simultaneously explain the right mass of left-handed (SM)  Majorana neutrino and KeV-scale mass of the right-handed (sterile) neutrino along with tiny mixing angle between them. 

 

\subsection{Forbidding Thermalization of KeV-Sterile Neutrino}
The production of sterile neutrinos obtained through the decay of $\phi$ can keep the same in thermal equilibrium with the plasma which might lead to $\Delta N_{\rm eff} \geq 1$.
The Feynman diagram of the decay channel is shown in fig.~6. 
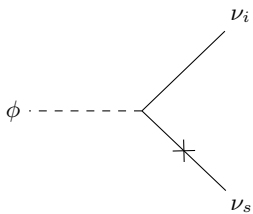
\begin{figure}[htbp]
\vskip 0.1in
\begin{center}
\begin{tikzpicture}
	\begin{feynman}
		\vertex  (a) ;
		\vertex [above right=of a] (b) {$\nu_i$};
		\vertex [below right=of a] (c) {$\nu_s$};
		\vertex [left=of a ] (d) {$\phi$};
		
		\diagram*{
			(b) -- [] (a);
			(c) -- [insertion = 0.5] (a); 
			(a) -- [scalar] (d);
		};	
	\end{feynman}
\end{tikzpicture}
\end{center}
    \caption{Feynman diagram representing the decay mode ${\phi\rightarrow \nu_i \nu_s}$.}
\end{figure}\\

In order to avoid this, the Yukawa coupling (and the mixing angle) has to be sufficiently small. In this subsection, We have analysed whether the tiny mixing angle considered in this work is sufficient to forbid the thermalization of right-handed neutrino.  The decay width of ${\phi\rightarrow \nu_i \nu_s}$ will be given by:
 \begin{equation}
 \Gamma_{\phi\rightarrow \nu_i \nu_s} \approx \frac{g^2_{\phi} y^2_N v^2_u}{m_{\phi}},
\end{equation}
where $y_N$  corresponds to the Yukawa coupling between left-handed neutrino and the right-handed antineutrino, $v_u = v \sin \beta$, with  $v =246$ GeV, and $\sin\beta \sim {\cal O}(1)$. If we compare the decay width to the expansion rate of the universe $H= T^2/M_p$, where $T = m_{\phi}$ for $m_{\phi} \geq T_{\rm dec-BBN}$(between 0.1-1 MeV), we get
\begin{equation}
\frac{\Gamma}{H}  \approx \frac{g^2_{\phi} y^2_N v^2_u M_{p}}{m^{3}_{\phi}}.
\end{equation}
By considering $\sin\theta = \frac{y_N v_u}{m_N}$, the expression of $\Gamma/H$ can be written in the following parameterized form
\begin{equation}
    \frac{\Gamma}{H} = 10 \times \Bigl[\frac{\sin^2{2\theta}}{10^{-14}}\Bigr] \Bigl[\frac{m_{\nu_s}}{\rm KeV}\Bigr]^2 \Bigl[\frac{\rm MeV}{m_{\phi}}\Bigr]^3 \Bigl[\frac{g_{\phi}}{0.1}\Bigr]^2.
    \end{equation}
    We can clearly see that the thermalization of the right-handed neutrino can be avoided if we choose the mixing angle $\sin^{2}{2\theta}\leq 10^{-14}$. Interestingly, we have naturally obtained this bound in our phenomenological model and it is also consistent with the relevant parameter space hatched in section~IV.
    



\section{Concluding Remarks}
The self-interacting neutrinos play an important role in alleviating the tension between the value of Hubble constant obtained by using recent CMB measurements and the local distance
ladder measurements. The strong self-interaction between neutrinos can keep the same in thermal equilibrium with each other until late times, thus affecting the free-streaming of neutrinos. The impact of changes in free-streaming behaviour of neutrinos can be analysed as a change in the phase shift as well as amplitude of baryon acoustic oscillations of CMB, which can lead to increase in the value of present day Hubble constant $H_0$.  It has been found that the fit to
CMB data prefers a specific range of the self-interaction
between neutrinos for all three flavors of neutrino, given by $G_{\rm eff}$ to be $(4.7^{+0.4}_{-0.6}{\rm~MeV})^{-2}$, dubbed as ``strongly" interacting region, and $(89^{+171}_{-61}{\rm~MeV})^{-2}$, dubbed as ``moderately” interacting neutrinos regime respectively.  Recently, the self-interacting neutrino have also been proposed to obtain the right amount of relic abundance of KeV-scale sterile neutrino while being safe from all X-Ray constraints. The standard Dodelson-Widrow mechanism allows the production of sterile neutrino DM for an appropriate values of active-sterile neutrino mixing angle and around KeV-scale mass. However, the entire parameter space gets ruled out from X-Ray constraints. Interestingly, the self-interaction between neutrinos offers a modified version of DW mechanism which can keep the most salient feature of the Dodelson-Widrow mechanism while making the parameter space free from X-Ray constraints.  In this work, we have analysed the corelation between the parameter space of self-interacting neutrino model required by Hubble tension and modified DW-based production mechanism of KeV-scale neutrino DM. We have calculated the relic abundance of KeV-sterile neutrino by taking a specific range
of the strength of self-interaction required to solve Hubble tension (except the range of $G_{\rm eff}$ ruled out by various laboratory and cosmological bounds). We have also obtained the $m_{\nu_s} - \sin^2{2\theta}$
parameter space for the preferred range of $G_{\rm eff}$. Our results clearly indicate that the entire parameter space (consistent with both Hubble tension solution and KeV-scale sterile neutrino DM) is free from X-Ray constraints. As the interaction between neutrinos is mediated through massive scalar particle, it is possible that the decay of massive scalar into sterile neutrino would keep the same in equilibrium with thermal plasma until the epoch of BBN, thus producing $\Delta N_{\rm eff} \ge 1$.  We have ensured that it can be avoided for the choice of mass and mixing angle considered in this work. Thus, our results show an interesting synergy between the  Hubble tension motivated self-interacting neutrino and KeV-sterile neutrino Dark Matter in the context of a consistent phenomenological model. If the discrepancy between the value of Hubble constant remains persistent even in future CMB observations, it would become an interesting testing ground to observe the existence of KeV-sterile neutrino as WDM Candidate. The WDM candidates are also considered as a popular choice impacting the global signal of 21-cm observations from current and future planned experiments to detect 21-cm hydrogen. Recently, the mass of sterile neutrinos obtained through DW mechanism has been constrained to be $m_{\nu_s}> 15$ KeV from the forecast study of 21 cm global signal from Square kilometer array (SKA)~\cite{Giri:2022nxq} and  $m_{\nu_s}> 15$ KeV from the observations of 21-cm global signal by EDGES collaboration~\cite{Vipp:2021obj} respectively. As a future direction, we would re-estimate the bounds on sterile neutrino DM from 21-cm global signal by considering self-interaction of neutrinos.

From theoretical model building point of view, we needed KeV-scale mass of sterile neutrino, very tiny mixing angle between sterile neutrino and SM neutrino, and the observed mass of SM neutrino. We have explained all the required values of mass and mixing parameters by embedding this scenario in a consistent phenomenological model obtained in the context of gravity mediated supersymmetic theory. Interestingly, our model could explain all the values needed to explain the consistency between the Hubble tension solution and KeV-scale sterile neutrino DM by pushing the supersymmetry breaking scale to be around ${\cal O}(10)$ TeV. Thus, the model has a potential to obtain the scale of supersymmetry from cosmological observations.
\section*{Acknowledgments}
 MD  would like to acknowledge support through DST-Inspire Faculty Fellowship of the Department of Science and Technology (DST), Government of India under the Grant Agreement number: IFA18-PH215. Both MD and AP would also like to thank IITRAM Ahmedabad where significant part of the work was done. AP is thankful to DST for the financial support provided under DST-Inspire Faculty Grant number: IFA18-PH215. MD is grateful to Gaurav Goswami for valuable suggestions and comments. AP is thankful to Chandan Kumar Sahu and Raj Kishore for their support with Mathematica programming.
\bibliography{bibliography}

\begin{thebibliography}{10}

\bibitem{DiValentino:2021izs}
Eleonora Di~Valentino, Olga Mena, Supriya Pan, Luca Visinelli, Weiqiang Yang,
  Alessandro Melchiorri, David~F. Mota, Adam~G. Riess, and Joseph Silk.
\newblock {In the realm of the Hubble tension\textemdash{}a review of
  solutions}.
\newblock {\em Class. Quant. Grav.}, 38(15):153001, 2021.

\bibitem{Abdalla:2022yfr}
Elcio Abdalla et~al.
\newblock {Cosmology intertwined: A review of the particle physics,
  astrophysics, and cosmology associated with the cosmological tensions and
  anomalies}.
\newblock {\em JHEAp}, 34:49--211, 2022.

\bibitem{Vagnozzi:2019ezj}
Sunny Vagnozzi.
\newblock {New physics in light of the $H_0$ tension: An alternative view}.
\newblock {\em Phys. Rev. D}, 102(2):023518, 2020.

\bibitem{Planck:2018vyg}
N.~Aghanim et~al.
\newblock {Planck 2018 results. VI. Cosmological parameters}.
\newblock {\em Astron. Astrophys.}, 641:A6, 2020.
\newblock [Erratum: Astron.Astrophys. 652, C4 (2021)].

\bibitem{Riess:2019cxk}
Adam~G. Riess, Stefano Casertano, Wenlong Yuan, Lucas~M. Macri, and Dan
  Scolnic.
\newblock {Large Magellanic Cloud Cepheid Standards Provide a 1\% Foundation
  for the Determination of the Hubble Constant and Stronger Evidence for
  Physics beyond $\Lambda$CDM}.
\newblock {\em Astrophys. J.}, 876(1):85, 2019.

\bibitem{riess2018new}
Adam~G Riess, Stefano Casertano, Wenlong Yuan, Lucas Macri, Jay Anderson,
  John~W MacKenty, J~Bradley Bowers, Kelsey~I Clubb, Alexei~V Filippenko,
  David~O Jones, et~al.
\newblock New parallaxes of galactic cepheids from spatially scanning the
  hubble space telescope: Implications for the hubble constant.
\newblock {\em The Astrophysical Journal}, 855(2):136, 2018.

\bibitem{Niedermann:2021ijp}
Florian Niedermann and Martin~S. Sloth.
\newblock {Hot new early dark energy: Towards a unified dark sector of
  neutrinos, dark energy and dark matter}.
\newblock {\em Phys. Lett. B}, 835:137555, 2022.

\bibitem{Niedermann:2021vgd}
Florian Niedermann and Martin~S. Sloth.
\newblock {Hot new early dark energy}.
\newblock {\em Phys. Rev. D}, 105(6):063509, 2022.

\bibitem{Rezazadeh:2022lsf}
K.~Rezazadeh, A.~Ashoorioon, and D.~Grin.
\newblock {Cascading Dark Energy}.
\newblock 8 2022.

\bibitem{Dainotti:2022bzg}
Maria~Giovanna Dainotti, Biagio De~Simone, Tiziano Schiavone, Giovanni Montani,
  Enrico Rinaldi, Gaetano Lambiase, Malgorzata Bogdan, and Sahil Ugale.
\newblock {On the Evolution of the Hubble Constant with the SNe Ia Pantheon
  Sample and Baryon Acoustic Oscillations: A Feasibility Study for
  GRB-Cosmology in 2030}.
\newblock {\em Galaxies}, 10(1):24, 2022.

\bibitem{Dainotti:2021pqg}
Maria~Giovanna Dainotti, Biagio De~Simone, Tiziano Schiavone, Giovanni Montani,
  Enrico Rinaldi, and Gaetano Lambiase.
\newblock {On the Hubble constant tension in the SNe Ia Pantheon sample}.
\newblock {\em Astrophys. J.}, 912(2):150, 2021.

\bibitem{Cyr-Racine:2013jua}
Francis-Yan Cyr-Racine and Kris Sigurdson.
\newblock {Limits on Neutrino-Neutrino Scattering in the Early Universe}.
\newblock {\em Phys. Rev. D}, 90(12):123533, 2014.

\bibitem{Lancaster:2017ksf}
Lachlan Lancaster, Francis-Yan Cyr-Racine, Lloyd Knox, and Zhen Pan.
\newblock {A tale of two modes: Neutrino free-streaming in the early universe}.
\newblock {\em JCAP}, 07:033, 2017.

\bibitem{Oldengott:2017fhy}
Isabel~M. Oldengott, Thomas Tram, Cornelius Rampf, and Yvonne Y.~Y. Wong.
\newblock {Interacting neutrinos in cosmology: exact description and
  constraints}.
\newblock {\em JCAP}, 11:027, 2017.

\bibitem{Huang:2017egl}
Guo-yuan Huang, Tommy Ohlsson, and Shun Zhou.
\newblock {Observational Constraints on Secret Neutrino Interactions from Big
  Bang Nucleosynthesis}.
\newblock {\em Phys. Rev. D}, 97(7):075009, 2018.

\bibitem{Forastieri:2019cuf}
Francesco Forastieri, Massimiliano Lattanzi, and Paolo Natoli.
\newblock {Cosmological constraints on neutrino self-interactions with a light
  mediator}.
\newblock {\em Phys. Rev. D}, 100(10):103526, 2019.

\bibitem{RoyChoudhury:2020dmd}
Shouvik Roy~Choudhury, Steen Hannestad, and Thomas Tram.
\newblock {Updated constraints on massive neutrino self-interactions from
  cosmology in light of the $H_0$ tension}.
\newblock {\em JCAP}, 03:084, 2021.

\bibitem{Das:2021guu}
Anirban Das and Subhajit Ghosh.
\newblock {Self-interacting neutrinos as a solution to the Hubble tension?}
\newblock {\em PoS}, EPS-HEP2021:124, 2022.

\bibitem{Blinov:2019gcj}
Nikita Blinov, Kevin~James Kelly, Gordan~Z Krnjaic, and Samuel~D McDermott.
\newblock {Constraining the Self-Interacting Neutrino Interpretation of the
  Hubble Tension}.
\newblock {\em Phys. Rev. Lett.}, 123(19):191102, 2019.

\bibitem{Lyu:2020lps}
Kun-Feng Lyu, Emmanuel Stamou, and Lian-Tao Wang.
\newblock {Self-interacting neutrinos: Solution to Hubble tension versus
  experimental constraints}.
\newblock {\em Phys. Rev. D}, 103(1):015004, 2021.

\bibitem{RoyChoudhury:2022rva}
Shouvik Roy~Choudhury, Steen Hannestad, and Thomas Tram.
\newblock {Massive neutrino self-interactions and inflation}.
\newblock {\em JCAP}, 10:018, 2022.

\bibitem{DeGouvea:2019wpf}
Andr\'e De~Gouv\^ea, Manibrata Sen, Walter Tangarife, and Yue Zhang.
\newblock {Dodelson-Widrow Mechanism in the Presence of Self-Interacting
  Neutrinos}.
\newblock {\em Phys. Rev. Lett.}, 124(8):081802, 2020.

\bibitem{Kelly:2020pcy}
Kevin~J. Kelly, Manibrata Sen, Walter Tangarife, and Yue Zhang.
\newblock {Origin of sterile neutrino dark matter via secret neutrino
  interactions with vector bosons}.
\newblock {\em Phys. Rev. D}, 101(11):115031, 2020.

\bibitem{Benso_2022}
Cristina Benso, Werner Rodejohann, Manibrata Sen, and Aaroodd~Ujjayini
  Ramachandran.
\newblock Sterile neutrino dark matter production in presence of nonstandard
  neutrino self-interactions: An {EFT} approach.
\newblock {\em Physical Review D}, 105(5), mar 2022.

\bibitem{Dodelson:1993je}
Scott Dodelson and Lawrence~M. Widrow.
\newblock {Sterile-neutrinos as dark matter}.
\newblock {\em Phys. Rev. Lett.}, 72:17--20, 1994.

\bibitem{Abazajian:2021zui}
Kevork~N. Abazajian.
\newblock {Neutrinos in Astrophysics and Cosmology: Theoretical Advanced Study
  Institute (TASI) 2020 Lectures}.
\newblock {\em PoS}, TASI2020:001, 2021.

\bibitem{Dekker:2021bos}
Ariane Dekker, Ebo Peerbooms, Fabian Zimmer, Kenny C.~Y. Ng, and Shin'ichiro
  Ando.
\newblock {Searches for sterile neutrinos and axionlike particles from the
  Galactic halo with eROSITA}.
\newblock {\em Phys. Rev. D}, 104(2):023021, 2021.

\bibitem{Tremaine:1979we}
S.~Tremaine and J.~E. Gunn.
\newblock {Dynamical Role of Light Neutral Leptons in Cosmology}.
\newblock {\em Phys. Rev. Lett.}, 42:407--410, 1979.

\bibitem{Boyarsky:2008ju}
Alexey Boyarsky, Oleg Ruchayskiy, and Dmytro Iakubovskyi.
\newblock {A Lower bound on the mass of Dark Matter particles}.
\newblock {\em JCAP}, 03:005, 2009.

\bibitem{Merle:2015vzu}
Alexander Merle, Aurel Schneider, and Maximilian Totzauer.
\newblock {Dodelson-Widrow Production of Sterile Neutrino Dark Matter with
  Non-Trivial Initial Abundance}.
\newblock {\em JCAP}, 04:003, 2016.

\bibitem{Abazajian:2017tcc}
Kevork~N. Abazajian.
\newblock {Sterile neutrinos in cosmology}.
\newblock {\em Phys. Rept.}, 711-712:1--28, 2017.

\bibitem{Watson:2011dw}
Casey~R. Watson, Zhi-Yuan Li, and Nicholas~K. Polley.
\newblock {Constraining Sterile Neutrino Warm Dark Matter with Chandra
  Observations of the Andromeda Galaxy}.
\newblock {\em JCAP}, 03:018, 2012.

\bibitem{Horiuchi:2013noa}
Shunsaku Horiuchi, Philip~J. Humphrey, Jose Onorbe, Kevork~N. Abazajian, Manoj
  Kaplinghat, and Shea Garrison-Kimmel.
\newblock {Sterile neutrino dark matter bounds from galaxies of the Local
  Group}.
\newblock {\em Phys. Rev. D}, 89(2):025017, 2014.

\bibitem{Perez:2016tcq}
Kerstin Perez, Kenny C.~Y. Ng, John~F. Beacom, Cora Hersh, Shunsaku Horiuchi,
  and Roman Krivonos.
\newblock {Almost closing the \ensuremath{\nu}MSM sterile neutrino dark matter
  window with NuSTAR}.
\newblock {\em Phys. Rev. D}, 95(12):123002, 2017.

\bibitem{Dessert:2018qih}
Christopher Dessert, Nicholas~L. Rodd, and Benjamin~R. Safdi.
\newblock {The dark matter interpretation of the 3.5-keV line is inconsistent
  with blank-sky observations}.
\newblock {\em Science}, 367(6485):1465--1467, 2020.

\bibitem{Ng:2019gch}
Kenny C.~Y. Ng, Brandon~M. Roach, Kerstin Perez, John~F. Beacom, Shunsaku
  Horiuchi, Roman Krivonos, and Daniel~R. Wik.
\newblock {New Constraints on Sterile Neutrino Dark Matter from $NuSTAR$ M31
  Observations}.
\newblock {\em Phys. Rev. D}, 99:083005, 2019.

\bibitem{calore2022constraints}
Francesca Calore, Ariane Dekker, Pasquale~Dario Serpico, and Thomas Siegert.
\newblock Constraints on light decaying dark matter candidates from 16 years of
  integral/spi observations.
\newblock {\em arXiv preprint arXiv:2209.06299}, 2022.

\bibitem{Benso_2019}
Cristina Benso, Vedran Brdar, Manfred Lindner, and Werner Rodejohann.
\newblock Prospects for finding sterile neutrino dark matter at {KATRIN}.
\newblock {\em Physical Review D}, 100(11), dec 2019.

\bibitem{Drewes:2016upu}
M.~Drewes et~al.
\newblock {A White Paper on keV Sterile Neutrino Dark Matter}.
\newblock {\em JCAP}, 01:025, 2017.

\bibitem{Viel:2013fqw}
Matteo Viel, George~D. Becker, James~S. Bolton, and Martin~G. Haehnelt.
\newblock {Warm dark matter as a solution to the small scale crisis: New
  constraints from high redshift Lyman-\ensuremath{\alpha} forest data}.
\newblock {\em Phys. Rev. D}, 88:043502, 2013.

\bibitem{Das:2020xke}
Anirban Das and Subhajit Ghosh.
\newblock {Flavor-specific interaction favors strong neutrino self-coupling in
  the early universe}.
\newblock {\em JCAP}, 07:038, 2021.

\bibitem{Ghosh:2019tab}
Subhajit Ghosh, Rishi Khatri, and Tuhin~S. Roy.
\newblock {Can dark neutrino interactions phase out the Hubble tension?}
\newblock {\em Phys. Rev. D}, 102(12):123544, 2020.

\bibitem{Shi:1998km}
Xiang-Dong Shi and George~M. Fuller.
\newblock {A New dark matter candidate: Nonthermal sterile neutrinos}.
\newblock {\em Phys. Rev. Lett.}, 82:2832--2835, 1999.

\bibitem{Shaposhnikov:2006xi}
Mikhail Shaposhnikov and Igor Tkachev.
\newblock {The nuMSM, inflation, and dark matter}.
\newblock {\em Phys. Lett. B}, 639:414--417, 2006.

\bibitem{Roland:2014vba}
Samuel~B. Roland, Bibhushan Shakya, and James~D. Wells.
\newblock {Neutrino Masses and Sterile Neutrino Dark Matter from the PeV
  Scale}.
\newblock {\em Phys. Rev. D}, 92(11):113009, 2015.

\bibitem{Asaka:2006ek}
Takehiko Asaka, Mikhail Shaposhnikov, and Alexander Kusenko.
\newblock {Opening a new window for warm dark matter}.
\newblock {\em Phys. Lett. B}, 638:401--406, 2006.

\bibitem{Kopp:2021jlk}
Joachim Kopp.
\newblock {Sterile neutrinos as dark matter candidates}.
\newblock {\em SciPost Phys. Lect. Notes}, 36:1, 2022.

\bibitem{Chichiri:2021wvw}
Carlos Chichiri, Graciela~B. Gelmini, Philip Lu, and Volodymyr Takhistov.
\newblock {Cosmological dependence of sterile neutrino dark matter with
  self-interacting neutrinos}.
\newblock {\em JCAP}, 09:036, 2022.

\bibitem{Lessa:2007up}
A.~P. Lessa and O.~L.~G. Peres.
\newblock {Revising limits on neutrino-Majoron couplings}.
\newblock {\em Phys. Rev. D}, 75:094001, 2007.

\bibitem{Esteban:2021tub}
Ivan Esteban, Sujata Pandey, Vedran Brdar, and John~F. Beacom.
\newblock {Probing secret interactions of astrophysical neutrinos in the
  high-statistics era}.
\newblock {\em Phys. Rev. D}, 104(12):123014, 2021.

\bibitem{roach2022long}
Brandon~M Roach, Steven Rossland, Kenny~CY Ng, Kerstin Perez, John~F Beacom,
  Brian~W Grefenstette, Shunsaku Horiuchi, Roman Krivonos, and Daniel~R Wik.
\newblock Long-exposure nustar constraints on decaying dark matter in the
  galactic halo.
\newblock {\em arXiv preprint arXiv:2207.04572}, 2022.

\bibitem{Boyarsky:2007ge}
Alexey Boyarsky, Denys Malyshev, Andrey Neronov, and Oleg Ruchayskiy.
\newblock {Constraining DM properties with SPI}.
\newblock {\em Mon. Not. Roy. Astron. Soc.}, 387:1345, 2008.

\bibitem{Calore:2022pks}
Francesca Calore, Ariane Dekker, Pasquale~Dario Serpico, and Thomas Siegert.
\newblock {Constraints on light decaying dark matter candidates from 16 years
  of INTEGRAL/SPI observations}.
\newblock 9 2022.

\bibitem{Sarkar:2008xir}
Utpal Sarkar.
\newblock {\em {Particle and astroparticle physics}}.
\newblock Series in high energy physics, cosmology, and gravitation. Taylor \&
  Francis, New York, USA, 2008.

\bibitem{Dhuria:2011ye}
Mansi Dhuria and Aalok Misra.
\newblock {Towards Large Volume Big Divisor D3-D7 'mu-Split Supersymmetry' and
  Ricci-Flat Swiss-Cheese Metrics, and Dimension-Six Neutrino Mass Operators}.
\newblock {\em Nucl. Phys. B}, 855:439--507, 2012.

\bibitem{Dhuria:2012bc}
Mansi Dhuria and Aalok Misra.
\newblock {(N)LSP Decays and Gravitino Dark Matter Relic Abundance in Big
  Divisor (nearly) SLagy D3/D7 mu-Split SUSY}.
\newblock {\em Nucl. Phys. B}, 867:636--748, 2013.

\bibitem{Dhuria:2013syh}
Mansi Dhuria and Aalok Misra.
\newblock {Sizable electron/neutron electric dipole moment in $D3/D7 \mu$-split
  supersymmetry}.
\newblock {\em Phys. Rev. D}, 90(8):085023, 2014.

\bibitem{Giudice:1988yz}
G.~F. Giudice and A.~Masiero.
\newblock {A Natural Solution to the mu Problem in Supergravity Theories}.
\newblock {\em Phys. Lett. B}, 206:480--484, 1988.

\bibitem{Arkani-Hamed:2000oup}
Nima Arkani-Hamed, Lawrence~J. Hall, Hitoshi Murayama, David Tucker-Smith, and
  Neal Weiner.
\newblock {Small neutrino masses from supersymmetry breaking}.
\newblock {\em Phys. Rev. D}, 64:115011, 2001.

\bibitem{Martin:1997ns}
Stephen~P. Martin.
\newblock {A Supersymmetry primer}.
\newblock {\em Adv. Ser. Direct. High Energy Phys.}, 18:1--98, 1998.

\bibitem{Giri:2022nxq}
Sambit~K. Giri and Aurel Schneider.
\newblock {Imprints of fermionic and bosonic mixed dark matter on the 21-cm
  signal at cosmic dawn}.
\newblock {\em Phys. Rev. D}, 105(8):083011, 2022.

\bibitem{Vipp:2021obj}
Venno Vipp, Andi Hektor, and Gert H\"utsi.
\newblock {Rapid onset of the 21-cm signal suggests a preferred mass range for
  dark matter particle}.
\newblock {\em Phys. Rev. D}, 103(12):123002, 2021.

\end{thebibliography}
\bibliographystyle{unsrt}

\end{document}